\documentclass[a4paper,11pt,onecolumn]{article}
\usepackage{amsmath,amsfonts,amssymb,graphicx}    
\usepackage[backend=biber, sorting=none, maxnames=3, minnames=1]{biblatex}
\renewbibmacro*{author}{
  \printnames{author}
  \setunit{\addspace}
  \iffieldundef{collaboration}
    {}
    {\mkbibparens{\printfield{collaboration}}}}

\usepackage{subfigure}   
\usepackage{bm}          
\usepackage{abstract}    
\usepackage{geometry}
\usepackage{float}
\usepackage{color,xcolor}
\usepackage{array}   
\usepackage{multirow}
\usepackage{indentfirst}
\usepackage{pdfpages}
\usepackage{hyperref}
\usepackage{pifont}
\usepackage{lineno}
\usepackage{siunitx}
\usepackage{comment}

\usepackage{diagbox}
\usepackage{physics}
\usepackage{booktabs}

\hypersetup{colorlinks=true,
					 linkcolor=blue,
					  anchorcolor=blue,
					  citecolor=blue}

\def\Hhe4{$^4_{\Lambda}$He}
\def\Hh4{$^4_{\Lambda}$H}
\def\HT{$^3_{\Lambda}$H}
\def\HF{$^4_{\Lambda}$H}

\def\HHeTr{$^4_{\Lambda}\rm{He} \rightarrow {}^{3}\rm{He} + \rm{p} + \pi^-$}
\def\snn3{$\sqrt{s_{\rm{NN}}}$ = 3 GeV}

\def\Hhe4{$^4_{\Lambda}\hbox{He}$}
\def\Hh4{$^4_{\Lambda}\hbox{H}$}

\begin{document}
\title{\quad\\[0.5cm]
Production of the \Hhe4 hypernucleus in Au+Au collisions at $\sqrt{s_{\rm{NN}}}$ = 3 GeV with STAR}
\author{The STAR Collaboration}
%\input{STAR_authors_article.tex}
%\author{}
\date{}
\maketitle
%\vspace{-2.0cm}
%\begin{center}
%{\bfseries STAR Collaboration}\par
%\vspace{0.6em}
%\begingroup
%\small
%\noindent\STARauthors\par
%\vspace{1.2em}
%\footnotesize
%\setlength{\parindent}{0pt}% Consistent left alignment for every affiliation
%\setlength{\parskip}{2pt}% Small separation between affiliations
%\noindent\STARaffiliations\par
%\endgroup
%\vspace{1em}

\begin{center}
\begin{abstract}
The STAR experiment reports the first measurement of the \Hhe4 hypernuclear yield as a function of rapidity and transverse momentum in 0--50\% central Au+Au collisions at $\sqrt{s_{\rm{NN}}} =$ 3 GeV. The \Hhe4 is reconstructed via its three-body decay channel, \HHeTr, with a statistical significance of about 9.5 standard deviations. The \Hhe4 yield as a function of rapidity is consistent with that of \Hh4, and the rapidity-dependent \Hhe4/\Hh4 yield ratio is consistent with the corresponding $^3$He/t ratio. All measurements, including the transverse-momentum spectra, are reasonably described by the Jet AA Microscopic Transport Model (JAM) with a coalescence afterburner, suggesting a coalescence-based formation scenario for hypernuclei at this energy. The canonical thermal model reproduces the observed yield ratios but overpredicts the absolute hypernuclear yields.
\end{abstract}
\end{center}

\newpage
\section{Introduction}
The hyperon-nucleon (YN) interaction is important for understanding baryon-baryon interactions under SU(3) symmetry~\cite{Machleidt:2011zz,Haidenbauer:2019boi}. Because hyperons have short lifetimes, producing a hyperon beam for dedicated hyperon–nucleon or hyperon–hyperon scattering experiments is extremely challenging. Hyperons (e.g., $\Lambda, ~\Xi$) can bind with normal nuclei to form so-called hypernuclei. By studying the properties of hypernuclei, such as their binding energies and lifetimes, the strength of the YN interaction can be constrained~\cite{RevModPhys.88.035004,Chen:2023mel}. Based on hypernuclear data, theoretical studies have predicted that the $\Lambda$ hyperon would appear in the cores of neutron stars~\cite{Schaffner-Bielich:2008zws}. However, the appearance of the $\Lambda$ hyperon would significantly soften the nuclear-matter equation of state (EoS), which contradicts observations of neutron stars with masses of about 2 solar masses. This is referred to as the ``hyperon puzzle"~\cite{Bombaci:2016xzl}. Proposed theoretical approaches for solving the hyperon puzzle include additional density-dependent YN interactions and three-body hyperon-nucleon-nucleon (YNN) interactions~\cite{Gerstung:2020ktv,Lonardoni:2014bwa}. Experimentally extracting the density-dependent YN interaction remains challenging. 

Heavy-ion collisions (HICs) are the only effective way to create dense nuclear matter in the laboratory. Nuclear matter with approximately 2--5 times the saturation density ($\rho_0$) can be created at a center-of-mass energy $\sqrt{s_{\rm{NN}}}$ of a few GeV~\cite{SORENSEN2024104080}. Hypernuclei production and their collective flow in HICs may enable investigation of the YN interaction in a nuclear medium with finite density and pressure~\cite{PhysRevLett.130.212301,Chen:2024aom}.
The coalescence mechanism~\cite{STEINHEIMER201285} for hypernuclei production in HICs assumes that nuclei and hyperons combine to form hypernuclei when they are close in both momentum and coordinate space at the freeze-out stage. In this work, Au+Au collisions at \snn3 were simulated with the Jet AA Microscopic Transport Model (JAM)~\cite{refId0, JAM2.1}, a microscopic hadronic transport model that describes the non-equilibrium evolution of heavy-ion collisions through hadronic propagation, binary collisions, and resonance and string excitation and decay.
Light nuclei and hypernuclei are subsequently formed using a coalescence afterburner, which combines protons, neutrons, and $\Lambda$ hyperons based on their proximity in coordinate and momentum space at freeze-out.
The statistical thermal model has also been used to describe hadron yields in relativistic HICs~\cite{Andronic:2017pug}, with only a few parameters, such as the freeze-out temperature, baryon chemical potential, and source volume.
The thermal model does not specify a dynamical formation mechanism for hypernuclei. Instead, it treats them as species whose abundances are determined by thermodynamic parameters and conserved-charge constraints at chemical freeze-out.
Therefore, particle yields predicted by the thermal model provide useful references for estimating yields and yield ratios. For HICs in the high-baryon-density region, the canonical ensemble treatment requires the inclusion of parameters related to strangeness production.

Yields of \HT \ and \HF \ have been measured in Au+Au collisions at \snn3 by the STAR Collaboration~\cite{PhysRevLett.128.202301}. 
Measuring production yields of additional hypernuclei (such as \Hhe4, $^4_{\Lambda\Lambda}$H) in HICs would be important for understanding the production mechanism. In a simple picture, a \Hhe4 (\HF) nucleus could be formed by combining $^3$He (t) with a $\Lambda$ hyperon; thus, comparing the yield ratios of \Hhe4/\HF \ and $^3$He/t could be used as an observable to probe the $^3$He-$\Lambda$ and t-$\Lambda$ interactions in the nuclear medium. 

In this paper, we report the first measurement of the \Hhe4 yield in Au+Au collisions at \snn3 through its three-body decay channel \HHeTr. The data were collected at the Relativistic Heavy Ion Collider (RHIC) by the Solenoidal Tracker At RHIC (STAR) experiment~\cite{ACKERMANN2003624} in 2018. Owing to limited statistics, the yield (d$N$/d$y$) and mean transverse momentum ($\langle p_{\rm{T}} \rangle$) of \Hhe4 as functions of rapidity are measured in a wide centrality bin of 0--50\%. The yield ratios of \Hhe4/\HF \ and $^3$He/t are extracted, and the results are compared with theoretical results from the JAM plus coalescence afterburner and the thermal model.

\section{Dataset and Event Selection}

These data were collected by the STAR experiment in 2018 using the fixed-target (FXT) configuration. In this configuration, a 0.25 mm gold foil target was installed 200.7 cm downstream of the STAR detector center, and a gold beam with an energy of 3.85 GeV/u (GeV per nucleon) was collided with the target. The corresponding center-of-mass rapidity is -1.045. The minimum-bias (MB) trigger condition is provided by the Beam-Beam Counters (BBC)~\cite{10.1063/1.2888113} and the Time of Flight (TOF) detector~\cite{Llope:2012zz}. It requires events to have at least one hit in the BBC on the downstream side of the fixed target and five or more hits in the TOF. To reduce background from interactions with the beam pipe, we require the radial position of the reconstructed collision vertex to be less than 1.5 cm from the beamline, which is well within the STAR beam pipe radius of 2 cm. Furthermore, we require the collision vertex position along the beam direction (z-axis) to lie within 2 cm of the target.

In this analysis, we use data from the most central collisions, corresponding to 0--50\% centrality. The collision centrality is determined by fitting the charged-particle multiplicity with a Glauber model~\cite{Miller_2007} in the pseudorapidity $\eta$ range $-2<\eta<0$. Details are given in Ref.~\cite{PhysRevC.103.034908}. To reduce pile-up contamination, events above the reference multiplicity of 195 are removed from the most central class~\cite{STARcentrality3GeV}. After the vertex cuts and centrality selection, approximately 160 million good events are selected for analysis.
\section[]{Particle identification}

\begin{figure}[htbp]
\centering
{\includegraphics[width=0.60\textwidth]{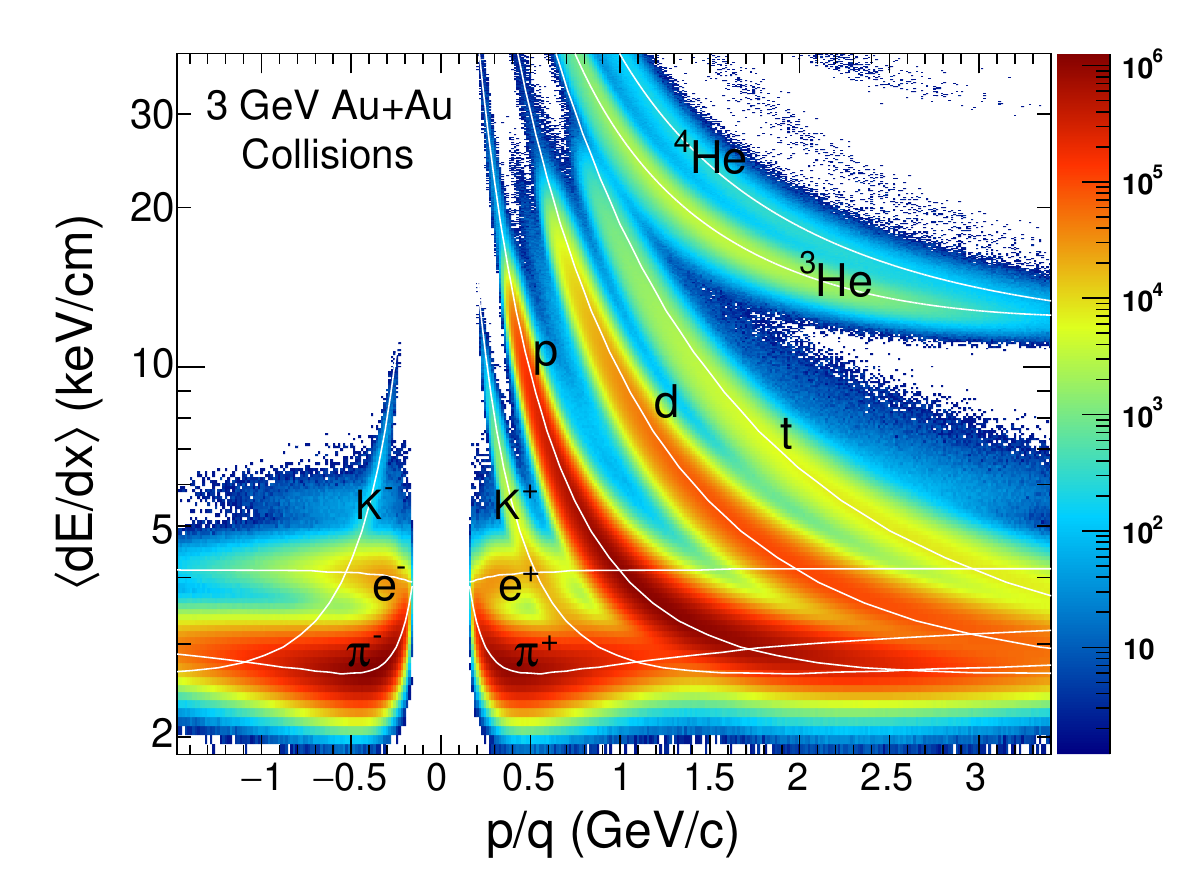}}
\caption{Rigidity ($p/q$) versus energy loss $\langle{\dv*{E}{x}}\rangle$ measured with the TPC detector in Au+Au collisions at $\sqrt{s_{\rm{NN}}}$ = 3.0 GeV; the lines represent the Bichsel functions for different particle species.}
\label{fig_PID}
\end{figure}

In this analysis, particle identification (PID) is performed with the Time Projection Chamber (TPC)~\cite{Anderson:2003ur} detector. The TPC is the main tracking detector in STAR, with a length of 4.2 m and a diameter of 4 m. Located within a large solenoidal magnet that operates in a 0.5 T magnetic field, the TPC contains a uniform electric field of $\approx\rm{135~V/cm}$, which is generated by a thin conductive central membrane, concentric field-cage cylinders, and the readout end caps. The average energy loss ($\langle{\dv*{E}{x}}\rangle$) in the TPC gas is a valuable tool for identifying particle species. It is averaged over measurements from up to 45 pad rows, depending on the track pseudorapidity. To avoid track splitting, at least 52\% of the maximum possible fit points are required for each track. Additionally, a minimum of 15 fit points is required to ensure the quality of track fitting. Figure \ref{fig_PID} shows the $\langle{\dv*{E}{x}}\rangle$ for particles in the TPC as a function of particle rigidity ($p/q$), where $p$ is the particle momentum and $q$ is the particle charge. The different bands represent different particle species, and the central values are predicted by the Bichsel function~\cite{BICHSEL2006154}. Given that the TPC provides reliable particle identification in the small $p/q$ region, protons are selected from the $\langle{\dv*{E}{x}}\rangle$ band within the $\mu \pm 3\sigma$ region for $p/q < 1.6$ GeV/$c$ and within the $\mu - 3\sigma < \langle{\dv*{E}{x}}\rangle < \mu + 2\sigma$ region for $p/q \geq$ 1.6 GeV/$c$ to suppress contamination from deuterons. Here, $\mu$ is the peak position of the $\langle{\dv*{E}{x}}\rangle$ band, and $\sigma$ is the Gaussian width. 
For $\pi^{-}$ identification, the $\langle{\dv*{E}{x}}\rangle$ band is well separated from those of other particle species in the low $p/q$ region, where most of the pions from the \Hhe4 decay are located. For $^{3}\rm{He}$, the corresponding $\langle{\dv*{E}{x}}\rangle$ band is well separated from those of other particle species over the full $p/q$ range used in this analysis. Therefore, a symmetric selection of $\mu \pm 3\sigma$ around the band centroid is applied throughout the full rigidity interval.

\begin{figure}[htbp]
\centering
{\includegraphics[width=0.9\textwidth]{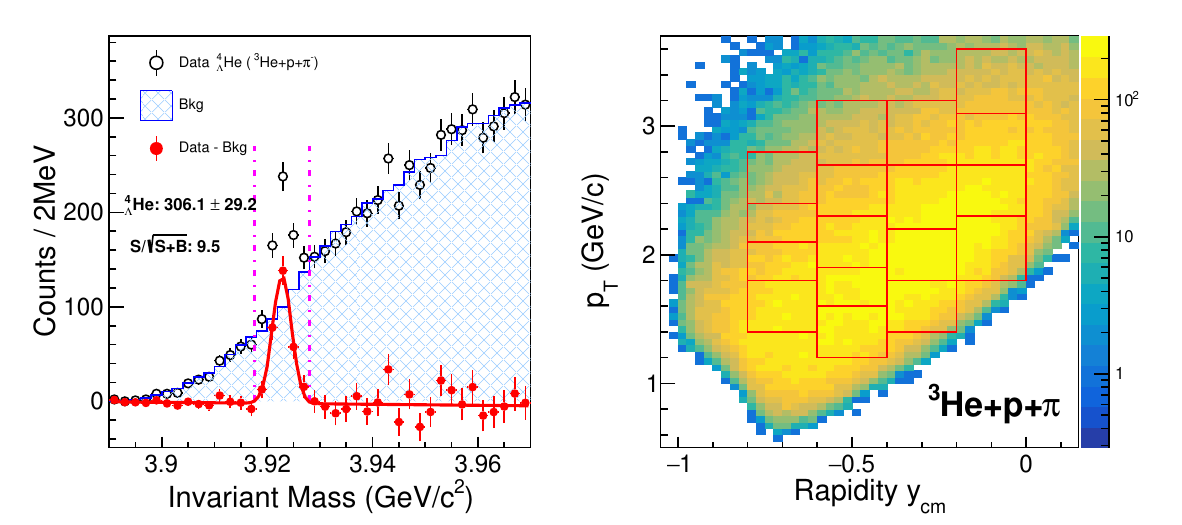}}
\caption{(Left panel) Invariant-mass distributions of $^3$He+p+$\pi^-$ from Au+Au collisions at $\sqrt{s_{\rm{NN}}}$ = 3.0 GeV. Open circles and filled red circles represent the raw data and background-subtracted distributions, respectively. The magenta vertical lines indicate the boundaries of the $\pm$3$\sigma$ signal region around the fitted invariant-mass peak. (Right panel) Acceptance of the \Hhe4 candidates as a function of transverse momentum $p_{\rm{T}}$ and rapidity $y$. The target rapidity in the center-of-mass frame is -1.045.}
\label{fig_topo}
\end{figure}

\Hhe4 is reconstructed via the three-body decay channel \HHeTr~with a branching ratio of 0.23~\cite{PhysRevC.76.039904,PhysRevD.6.3069}. Based on the Kalman Filter (KF) algorithm~\cite{Zyzak2016}, the KF particle finder package~\cite{Kisel_2020} is used to reconstruct and select short-lived particles. It performs Kalman-filter fits of tracks and vertices and provides $\chi^2$ values that characterize the quality of each topological constraint. Specifically, $\chi^{2}_{\rm{topo}}$ is the $\chi^2$ associated with constraining the reconstructed mother-particle trajectory to originate from the primary vertex, and therefore quantifies how well the reconstructed \Hhe4 candidate points back to the collision vertex. $\chi^{2}_{\rm{ndf}}$ is the $\chi^2/\rm{NDF}$ of the common-vertex fit of the three daughter tracks, reflecting the probability that the daughter trajectories originate from the same decay vertex. $\chi^{2}_{\rm{prim}}$ is the $\chi^2$ of an individual daughter track with respect to the primary vertex and is used to discriminate decay daughters (large $\chi^{2}_{\rm{prim}}$) from primary tracks (small $\chi^{2}_{\rm{prim}}$). We require $\chi^{2}_{\rm{topo}}$ to be less than 3.0, $\chi^{2}_{\rm{ndf}}$ to be less than 5.0, and $\chi^{2}_{\rm{prim}}$ for the proton or pion to be larger than 15.0. The default topological selection criteria are determined by scanning different combinations of cut values and choosing those that optimize the statistical significance of the reconstructed signal while retaining sufficient signal counts. Additionally, we require the decay length to be larger than 8.0~cm. To reconstruct the combinatorial background of \Hhe4 candidates, we rotate the $^3$He tracks in the transverse plane around the primary vertex by a random angle between 10 and 350 degrees to break kinematic correlations, and then repeat the analysis using the same reconstruction procedure and topological cuts. To improve the background statistics and smoothness, we repeat this rotation 20 times. Figure~\ref{fig_topo} (left) shows the invariant mass spectrum of \Hhe4 candidates and the rotated background. 
The rotated background is normalized to the data in the high-mass sideband, $3.94 <M< 3.97~\mathrm{GeV/c^{2}}$, which lies more than 6$\sigma$ from the signal peak. The low-mass sideband is not used because it is close to the kinematic limit.
The spectrum is fitted with a Gaussian function plus a $\rm{1^{st}}$-order polynomial, resulting in a total yield of over 300 counts and a statistical significance of approximately 10. Figure~\ref{fig_topo} (right) shows the acceptance of the raw \Hhe4 candidates in the transverse-momentum--rapidity ($p_{\rm{T}}$--$y$) plane; $y$ is defined in the center-of-mass frame. The red boxes indicate the binning scheme used in the following differential-yield analysis. The acceptance is mainly constrained by the TPC pseudo-rapidity ($\eta$) coverage and the lower-limit cuts on particle $p_{\rm{T}}$.

\section[]{Reconstruction efficiency}
To obtain the final yield spectra in each rapidity interval, PID and reconstruction efficiency corrections must be applied. As mentioned above, particles are selected from the $\langle{\dv*{E}{x}}\rangle$ band using different widths. The PID efficiency can be corrected directly using the inverse of the Gaussian integral probability within different $p/q$ ranges. The reconstruction efficiency is estimated using an embedding method, in which Monte Carlo (MC) simulations of \Hhe4 decays are embedded into MB events at a 5\% ratio and then processed through the same reconstruction procedure with the same topological cuts. The detector responses to MC tracks are simulated using GEANT3~\cite{Brun:1987ma}. The ratio of reconstructed MC tracks to embedded MC tracks gives the reconstruction efficiency.

Because the embedded signal cannot accurately reproduce some distributions of the real signal, several kinematic distributions are re-weighted to obtain a reliable efficiency estimate. In this analysis, the proper decay length distribution of the embedded signal is re-weighted so that its lifetime corresponds to the world-average value of 245 ps~\cite{PhysRevC.76.035501}. The $p_{\rm{T}}$ spectrum of the embedded signal is further re-weighted to match a negative exponential distribution. Finally, a Dalitz-plot re-weight is applied to reproduce the reconstructed Dalitz distribution of $M_{p\pi}$ vs. $M_{p\rm{{^3}He}}$ obtained from data, where $M_{p\pi}$ and $M_{p\rm{{^3}He}}$ represent the invariant masses of $p+\pi$ and $p+\rm{{^3}He}$, respectively. These corrections modify the phase-space distributions of the daughter particles and therefore affect the reconstruction efficiency. The $p_{\rm{T}}$ spectrum and Dalitz-plot re-weights are not fully independent because both modify the kinematic phase-space distributions of the decay daughters; consequently, applying one re-weight can modify the distribution used to determine the other. To account for this effect, we alternatively include the $p_{\rm{T}}$ spectrum re-weighting when determining the Dalitz-plot correction and vice versa. The resulting difference is propagated through the analysis and assigned as the systematic uncertainty associated with the correlation between the two re-weights.

The relative impact of each re-weight depends on the topological selections. In this analysis, the Dalitz-plot re-weight leads to an average change of about 20\% in the reconstruction efficiency, while the $p_{\rm{T}}$ spectrum and proper-decay-length re-weights modify the efficiency by less than 10\% and approximately 10\%, respectively.

Since GEANT3 treats light nuclei as generic hadrons, the absorption of $\rm{{^3}He}$ by the detector material is not properly modeled in the embedding simulation. The correction is obtained from dedicated GEANT4 simulations using the STAR material geometry, in which the $p_{\rm{T}}$-dependent non-absorption efficiencies of protons and light nuclei are evaluated. Their ratio is then used to correct the GEANT3 embedding efficiency~\cite{PhysRevLett.130.202301}.

\begin{figure}[htbp]
	\centering
	{\includegraphics[width=0.6\textwidth]{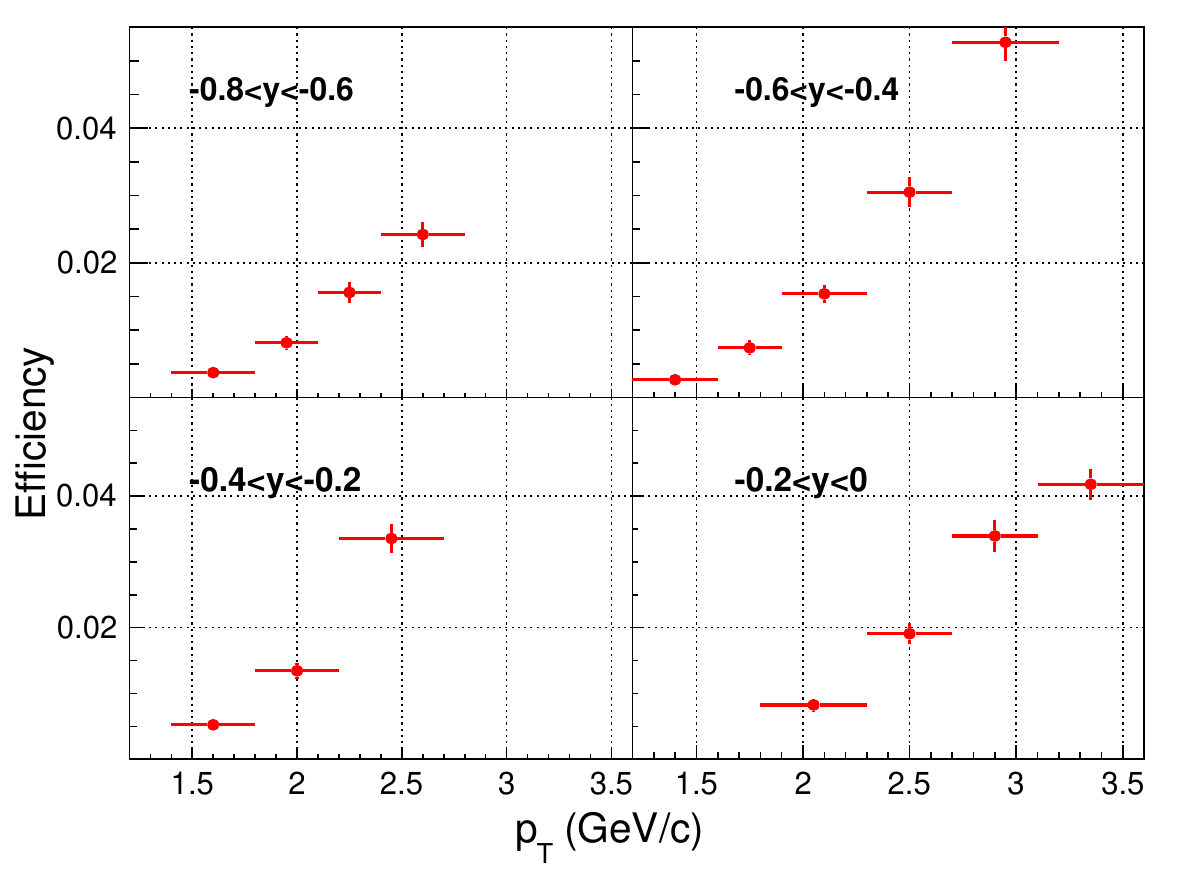}}
	\caption{Efficiency of \Hhe4 reconstruction as a function of $p_{\rm{T}}$ in different rapidity regions.}
	\label{efficiency_4raps}
\end{figure}

Taking into account the above re-weights and corrections, the efficiencies of \Hhe4 reconstruction versus $p_{\rm{T}}$ in different rapidity regions are shown in Figure~\ref{efficiency_4raps}.

\section[]{Transverse-momentum spectra and extrapolation method}

The differential yields of \Hhe4 as functions of $p_{\rm{T}}$ in different rapidity bins are calculated using Eq. \eqref{ptSpectrum_calculaton}.
\begin{equation} \label{ptSpectrum_calculaton}
\frac{d^{2}N}{d p_{\rm{T}}dy}=\frac{N_{\mathrm{raw}}}{N_{\mathrm{evt}}\Delta p_{\rm{T}}\Delta y}\times\frac{1}{\varepsilon}\times\frac{1}{B.R.({}^4_\Lambda \mathrm{He} \rightarrow {}^{3}\mathrm{He} + p + \pi^{-})}
\end{equation}
Here, $\rm N_{evt}$ denotes the total number of events in the 0--50\% centrality interval, $\rm N_{raw}$ is the raw yield extracted in each bin, and $\Delta p_{\rm{T}}$ and $\Delta y$ are the corresponding bin widths. The binning scheme is shown in Figure~\ref{fig_topo} (right). $\varepsilon$ is the corresponding mean efficiency, and $B.R.$ is the branching ratio for the measured decay channel. Several models can be used to describe the $p_{\rm{T}}$ spectrum. In this analysis, the blast-wave model~\cite{Schnedermann_1993} is used to describe the $p_{\rm{T}}$ spectrum and to extrapolate to the low- and high-$p_{\rm{T}}$ regions, as expressed by equation \eqref{eq_blastWave}, which gives the invariant differential yield: 
\begin{equation} \label{eq_blastWave}
\frac{1}{2\pi p_{\rm{T}}}\frac{d^{2}N}{d p_{\rm{T}}d y}\propto\int_{0}^{R}r d r m_{\rm{T}}I_{0}\left(\frac{p_{\rm{T}}\sinh\rho(r)}{T_{k i n}}\right) \times K_{1}\left(\frac{m_{\rm{T}}\cosh \rho(r)}{T_{k i n}}\right),
\end{equation}
where $m_{\rm{T}}$ is the transverse mass of the particle, $I_\mathrm{0}$ and $K_\mathrm{1}$ are the modified Bessel functions, $T_{kin}$ is the kinetic freeze-out temperature, and $R$ denotes the transverse radius of the particle-emitting source at kinetic freeze-out. The transverse boost rapidity is given by $\rho(r)=\rm{tanh}^{-1}\beta_{T}$. The transverse radial flow velocity $\beta_{T}$ in the region $0 \leq r \leq R$ can be expressed as $\beta_{T} = \beta_{S} (r/R)^n$, where $\beta_{S}$ is the surface velocity and $n$ characterizes the shape of the flow velocity profile (fixed at $n = 1$ in this analysis).

\begin{figure}[htbp]
\centering
{\includegraphics[width=0.6\textwidth]{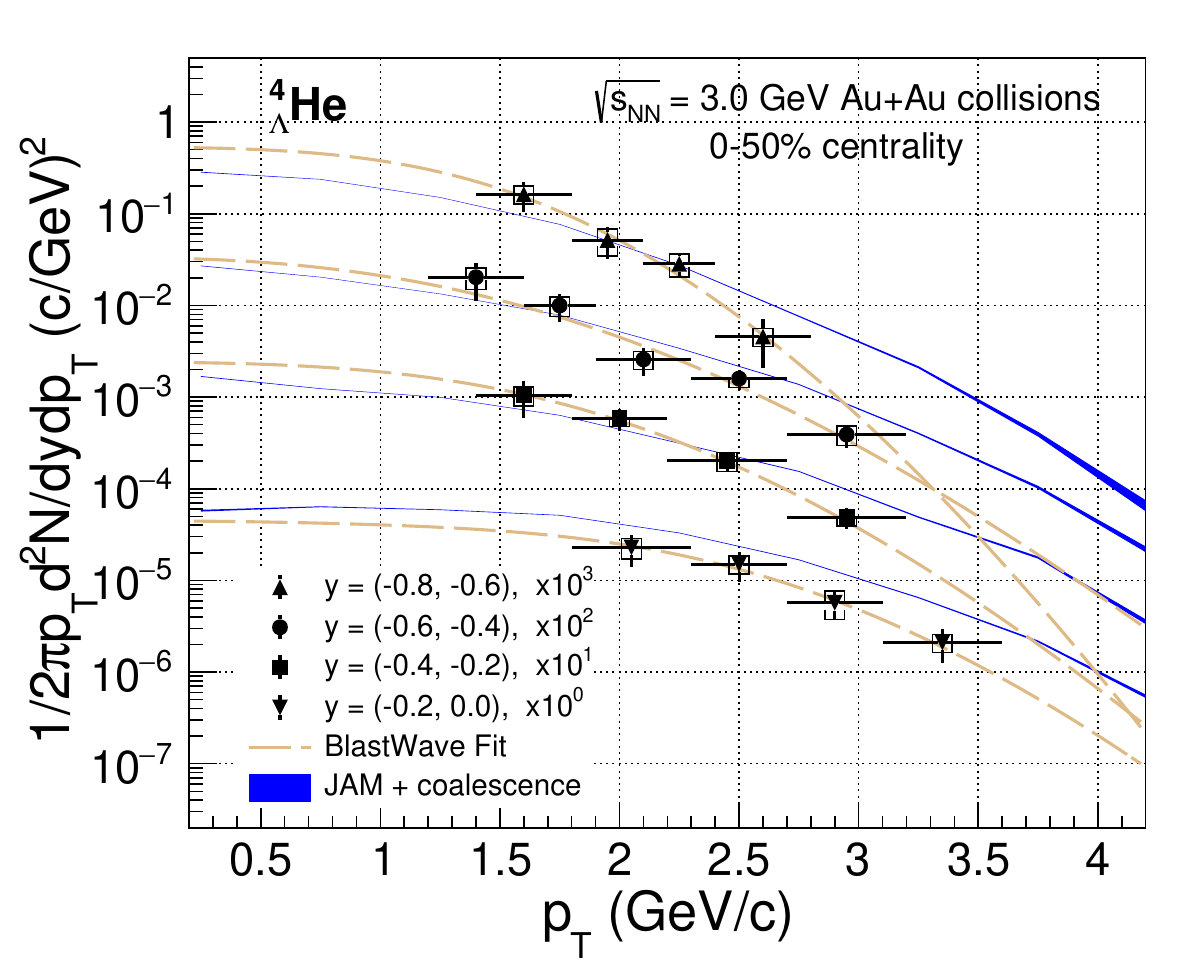}}
\caption{The invariant $p_{\rm{T}}$ spectra of \Hhe4 in $0$--$50\%$ central Au+Au collisions at $\sqrt{s_{\rm{NN}}}=3.0$ GeV are shown for four rapidity bins. The blast-wave fit results are shown as yellow dashed lines, and the JAM model results are shown as blue bands. Vertical bars and brackets represent statistical and systematic uncertainties, respectively.}
\label{fig_result_pt}
\end{figure}

Figure~\ref{fig_result_pt} shows the invariant yields as a function of $p_{\rm{T}}$ for \Hhe4 in rapidity intervals from the target-rapidity region ($-0.8 < y < -0.6$) to the midrapidity region ($-0.2 < y < 0$). The blast-wave fit results are shown as yellow dashed lines. For comparison, predictions from JAM with a coalescence afterburner are shown as blue bands. The model predictions are consistent with the experimental data, except in the midrapidity region, where the data points are systematically lower.

A spherical, sharp phase-space coalescence criterion is employed: a (hyper-)nucleus is formed when the relative spatial and momentum separations of its constituents satisfy $\Delta r < \Delta r_{\rm max}$ and $\Delta p < \Delta p_{\rm max}$, respectively. No explicit nuclear wave function or Wigner-function weighting is included in this prescription. For the $A=4$ hypernuclei, $\Delta r_{\rm max}=4.8$ fm and $\Delta p_{\rm max}=0.38$ GeV/$c$ are used.

These parameters are primarily constrained by the \Hh4 d$N$/d$y$ spectra~\cite{PhysRevLett.128.202301} because of their higher statistical precision. Using the fitted blast-wave function, d$N$/d$y$ and $\langle p_{\rm{T}} \rangle$ are extracted in different rapidity regions. The systematic uncertainties of the measured $p_{\rm{T}}$ spectra are evaluated in different rapidity intervals by varying the signal-extraction method and topological selections. Because these variations can introduce correlated changes among $p_{\rm{T}}$ bins, we do not propagate the point-by-point uncertainties of the $p_\mathrm{T}$ spectra to estimate the systematic uncertainties of d$N$/d$y$ and $\langle p_{\rm{T}} \rangle$. Instead, these integrated quantities are re-extracted for each analysis variation, and their deviations from the default results are evaluated, as detailed in Sec. 6. 

\section[]{Systematic uncertainties}
In this analysis, systematic uncertainties from different sources are considered independently for each rapidity region. One of the main sources is the variation of the topological-variable cuts. The distributions of the topological variables used in the analysis may not be perfectly reproduced by embedding. To estimate this systematic uncertainty, the topological cuts are varied, and the full analysis is repeated using different combinations of these varied cuts. The cut thresholds are tightened and loosened around their default values, typically by about 10--20\%. The root mean square values of the resulting d$N$/d$y$ and $\langle p_{\rm{T}} \rangle$ distributions are assigned as the systematic uncertainties from the topological-variable cuts. The choice of extrapolation function is another dominant source of systematic uncertainty for both d$N$/d$y$ and $\langle p_{\rm{T}} \rangle$. A blast-wave function is used by default to fit the $p_{\rm{T}}$ spectrum and extrapolate it to the low-$p_{\rm{T}}$ region. Alternatively, Boltzmann and $m_{\rm{T}}$ exponential functions are used to fit the $p_{\rm{T}}$ spectrum, as shown in equations \eqref{eq_mTExponential} and \eqref{eq_Boltzmann}, respectively:
\begin{equation} \label{eq_mTExponential}
{\rm Boltzmann} : \propto {\rm exp}(-m_{\rm{T}}/T_{\rm{B}}) 
\end{equation}
\begin{equation} \label{eq_Boltzmann}
m_{\rm{T}}~{\rm exponential} : \propto m_{\rm{T}}{\rm exp}(-m_{\rm{T}}/T_{\rm{B}}) 
\end{equation}
The systematic uncertainty due to the limited low-$p_{\rm{T}}$ acceptance is taken as half of the maximum absolute difference among the fit functions in the extrapolation region. Systematic uncertainties from other sources are also evaluated, including:
\begin{itemize}
    \item Tracking efficiency: Following standard STAR practice, a 5\% uncertainty is assigned to the reconstruction efficiency of each daughter track and propagated to the parent-particle yield. The uncertainties from different daughter tracks are conservatively treated as fully correlated and are therefore added linearly, resulting in relative uncertainties of 15\% for the \Hhe4 yield measurements. This uncertainty is not assigned to the $\langle p_T\rangle$ measurements because it primarily affects the overall normalization and largely cancels in $\langle p_T\rangle$. 
    \item Track-quality selection: The quality of single-track reconstruction can be affected by the requirement for the minimum number of TPC fit points. The default requirement of at least 15 fit points is increased to 18 to assess possible reconstruction biases. The resulting change in the efficiency-corrected yields is taken as the systematic uncertainty associated with the track-quality selection. 
    \item Signal extraction: By default, we fit the raw invariant mass spectrum in each $p_{\rm{T}} - y$ bin and use the Gaussian area parameter as the raw yield. Alternatively, we use a histogram bin-counting method, in which the total bin counts in the signal region are taken as the raw yield. In both approaches, efficiency corrections are then applied to obtain the corrected signal yield. This source quantifies the systematic uncertainty associated with possible deviations of the invariant mass spectrum from an ideal Gaussian shape.
    \item Background estimation: As the default approach, we use a first-order polynomial to estimate the residual background after rotation background subtraction. Alternatively, we fit the residual background with a constant term. The difference between these two approaches reflects the systematic uncertainty associated with a possible mismatch between the foreground and background invariant mass spectra in the sideband region. 
    \item Reweighting correlation: The correlation between the $p_{\rm{T}}$ spectrum and the Dalitz-plot reweighting factors is evaluated using the procedure described in Sec. 4. The resulting variations in d$N$/d$y$ and $\langle p_{\rm{T}} \rangle$ are assigned as the corresponding systematic uncertainties. 
\end{itemize}
Tables~\ref{dNdy_sysErr_all} and~\ref{meanPt_sysErr_all} summarize the systematic uncertainties in d$N$/d$y$ and $\langle p_{\rm{T}} \rangle$ from various sources in different rapidity regions. The total systematic uncertainty is obtained by adding the individual contributions in quadrature. Among all sources, the topological cuts and the extrapolation procedure dominate the overall systematic uncertainty for both d$N$/d$y$ and $\langle p_{\rm{T}} \rangle$. Their larger contributions in the target rapidity region are associated with the limited low-$p_{\rm{T}}$ coverage and the steeper $p_{\rm{T}}$ spectrum. These features increase the unmeasured fraction of the yield and, consequently, the model dependence of the extrapolated contribution, while also amplifying the impact of topological-cut variations on the integrated yields.

The systematic uncertainties in $\langle p_{\rm{T}} \rangle$ are generally smaller than those in d$N$/d$y$ because $\langle p_{\rm{T}} \rangle$ is calculated as the ratio of the first moment of the $p_\mathrm{T}$ spectrum to its integral. Systematic variations that mainly affect the overall normalization largely cancel in this ratio.

\begin{table}[htbp]
	\centering
	\caption{Systematic uncertainties in d$N$/d$y$ from all sources in different rapidity regions.}
	\label{dNdy_sysErr_all}
	\begingroup
	\setlength{\tabcolsep}{2pt}
	\begin{tabular}{lcccc}
		\toprule
		 & \multicolumn{4}{c}{Systematic uncertainty (\%)} \\
		\cmidrule(lr){2-5}
		& $-0.8<y<-0.6$ & $-0.6<y<-0.4$ & $-0.4<y<-0.2$ & $-0.2<y<0.0$ \\
		\midrule
		Topo-variable cuts   & 28.6 & 16.4 & 13.2 & 14.8 \\
		Extrapolation        & 33.7 &  7.0 & 18.6 & 27.2 \\
		Track-quality selection  &  4.7 &  2.3 &  2.4 &  3.7 \\
		Signal extraction    &  1.1 &  2.6 &  7.7 & 15.4 \\
		Background estimation&  1.0 &  3.4 &  5.3 &  1.6 \\
		Weight correlation   &  6.1 &  3.7 &  2.2 &  2.5 \\
		\midrule
		\textbf{Total}       & \textbf{47.3} & \textbf{24.1} & \textbf{29.0} & \textbf{38.0} \\
		\bottomrule
	\end{tabular}
	\endgroup
\end{table}

\begin{table}[htbp]
	\centering
	\caption{Systematic uncertainties in $\langle p_T \rangle$ from all sources in different rapidity regions.}
	\label{meanPt_sysErr_all}
	\begingroup
	\setlength{\tabcolsep}{2pt}
	\begin{tabular}{lcccc}
		\toprule
			& \multicolumn{4}{c}{Systematic uncertainty (\%)} \\
		\cmidrule(lr){2-5}
		& $-0.8<y<-0.6$ & $-0.6<y<-0.4$ & $-0.4<y<-0.2$ & $-0.2<y<0.0$ \\
		\midrule
		Topo-variable cuts   & 4.1 & 5.3 & 5.4 & 2.4 \\
		Extrapolation        & 8.0 & 2.6 & 6.0 & 7.8 \\
		Track-quality selection & 0.1 & 1.9 & 1.0 & 1.5 \\
		Signal extraction    & 0.6 & 1.9 & 2.8 & 4.6 \\
		Background estimation& 0.6 & 0.8 & 1.2 & 0.5 \\
		Weight correlation   & 0.5 & 0.7 & 0.7 & 0.2 \\
		\midrule
		\textbf{Total}       & \textbf{9.0} & \textbf{6.6} & \textbf{8.7} & \textbf{9.5} \\
		\bottomrule
	\end{tabular}
	\endgroup
\end{table}

\section[]{Results and discussion}

\begin{figure}[htbp] 
\centering
{\includegraphics[width=1.0\textwidth]{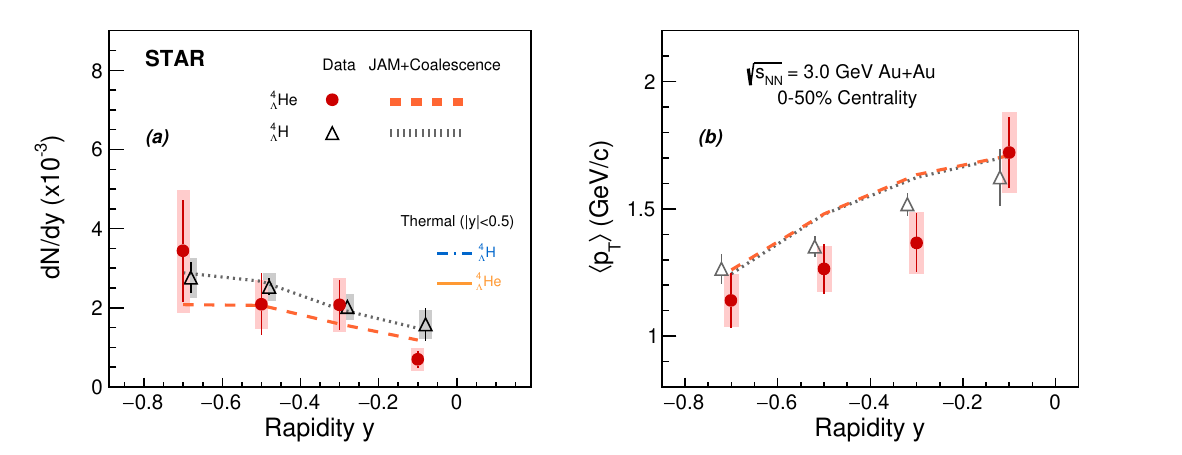}}
\caption{(a) Rapidity dependence of the \Hhe4 yields in 0--50\% central Au+Au collisions at $\sqrt{s_{\rm{NN}}}$=3.0 GeV. (b) Rapidity dependence of the mean transverse momentum, $\langle p_{\rm{T}} \rangle$, of \Hhe4. The \Hhe4 and \Hh4 data are shown as filled circles and open triangles, respectively, while the JAM+coalescence calculations are represented by dashed and dotted lines. Vertical bars and shaded boxes represent statistical and systematic uncertainties, respectively. The horizontal lines in panel (a) indicate the Thermal-FIST predictions for the d$N$/d$y$ of \Hhe4 and \Hh4 within $\vert y \vert < 0.5$.}
\label{fig_result_dNdy_meanPt}
\end{figure}

Figure~\ref{fig_result_dNdy_meanPt} shows the rapidity dependences of d$N$/d$y$ and $\langle p_{\rm{T}} \rangle$ for \Hhe4 measured in this work. The \Hh4 results derived from the previously published STAR measurement in Ref.~\cite{PhysRevLett.128.202301} are also shown for comparison. For consistency, the \Hh4 $p_{\rm{T}}$ spectra are refitted using the same blast-wave function as for \Hhe4, instead of the $m_T$-exponential function used in Ref.~\cite{PhysRevLett.128.202301}. The 0--10\% and 10--50\% results for \Hh4 are then combined to obtain the corresponding 0--50\% centrality interval, with the statistical and systematic uncertainties propagated accordingly. This enables a direct comparison of the two $A=4$ mirror hypernuclei using a common centrality interval and spectral treatment.

The model results from the JAM model coupled to a coalescence afterburner for \Hhe4 and \Hh4 are shown by the red dashed line and black dotted line, respectively. Here, the bootstrap method~\cite{10.1214/aos/1176344552} is applied to estimate the statistical uncertainties associated with the extrapolation region. This method is a widely used statistical resampling technique for estimating uncertainties when analytic error propagation is difficult to compute. Branching ratios of the \Hhe4 three-body and \Hh4 two-body decay channels are used to correct the yields. For \Hhe4, $\Gamma_{\pi^{-}}/\Gamma_{\rm{total}}=0.269\pm0.022\pm0.014$~\cite{PhysRevC.76.039904} and $\Gamma_{^4_{\Lambda}\rm{He}\rightarrow^3\rm{He}+p+\pi^-}/\Gamma_{\pi^{-}}=0.87\pm0.067$~\cite{PhysRevD.6.3069} give $\Gamma_{^4_{\Lambda}\rm{He}\rightarrow^3\rm{He}+p+\pi^-}/\Gamma_{\rm{total}}=0.23\pm0.029$. Here, $\Gamma$ denotes the decay width, and the subscripts indicate the corresponding decay channels. For \Hh4, the branching-ratio correction follows the treatment in Ref.~\cite{PhysRevLett.128.202301}, with $\Gamma_{^4_\Lambda\rm{H}\rightarrow^4\rm{He}+\pi^-}/\Gamma_{\rm{total}}=0.50\pm0.05$~\cite{OUTA1998251c}.

The branching-ratio uncertainties are treated as independent sources of systematic uncertainty. They are added in quadrature with the other contributions listed in Table~\ref{dNdy_sysErr_all} and included in the systematic uncertainties shown in Fig.~\ref{fig_result_dNdy_meanPt}.

It is found that, for both \Hhe4 and \Hh4, the d$N$/d$y$ values decrease from target rapidity to midrapidity, following the monotonic decrease of light-nuclei d$N$/d$y$ in the same collision system~\cite{PhysRevC.110.054911}. Similar to light nuclei in the same collision system~\cite{PhysRevC.110.054911}, both \Hhe4 and \Hh4 show a monotonic increase in $\langle p_{\rm{T}} \rangle$ from target rapidity to midrapidity.

With the same coalescence parameters used for \Hhe4 and \Hh4, the JAM plus coalescence calculations reasonably describe the measured d$N$/d$y$ distributions. The predicted $\langle p_{\rm{T}} \rangle$ values tend to lie above the data: the differences are within 2$\sigma$ for \Hhe4 at every measured rapidity, but reach approximately 3$\sigma$ for \Hh4 in the rapidity interval ($-0.6 < y < -0.4$).

The quoted deviations are obtained by dividing the absolute model--data differences by the experimental uncertainties, combining statistical and systematic uncertainties in quadrature for \Hhe4 and using statistical uncertainties alone for \Hh4. No model uncertainty is included here.

Statistical thermal model calculations are used to characterize the yields with the Thermal-FIST package~\cite{VOVCHENKO2019295} using the strangeness-canonical ensemble (CE). The chemical freeze-out temperature $T_{\rm ch}$ ($\approx 85$ MeV) and baryon chemical potential $\mu_{\rm B}$ ($\approx 728$ MeV) are taken from the parameterization of Ref.~\cite{Vovchenko:2015idt}, which reasonably describes the measured yields of pions, kaons, protons, and hyperons in central heavy-ion collisions. The strangeness correlation radius $r_c = 3.55$ fm and saturation factor $\gamma_{s}=1$ are used to describe strangeness production at this energy, including $K^0_{\rm{S}}$, $\Lambda$, and $\Xi^{-}$~\cite{2022137152,Abdulhamid2024}. The radius of the spherical thermal source, $R = 8.28$ fm for $\vert y \vert < 0.5$, is obtained from a fit to the charged-pion yields~\cite{PhysRevC.107.014912}. To obtain the predictions for 0--50\% centrality, the hypernuclear yields are scaled according to the $\langle N_{\rm{part}} \rangle$ dependence of strange-hadron production measured at $\sqrt{s_{\rm{NN}}}$=3.0 GeV~\cite{Abdulhamid2024}, assuming the same centrality dependence for hypernuclei~\cite{QMTalk_ZhouYingjie}. Contributions from excited (hyper-)nuclei and feed-down to \Hhe4 and \HF \ are also included. The predicted \Hhe4 and \Hh4 yields at midrapidity, $\vert y \vert < 0.5$, are shown as the yellow line and blue dot-dashed line in Fig.~\ref{fig_result_dNdy_meanPt}, respectively; both are higher than the experimental measurements.

Yield ratios provide a sensitive probe of the production mechanism in heavy-ion collisions. Combining the present \Hhe4 measurement with the previously published \Hh4 results enables a comparison with the corresponding light-nuclei ratio $^3$He/$t$. In a simple coalescence picture, \Hhe4\ (\HF) can be formed predominantly by coalescing a $\Lambda$ hyperon with a $^3$He\ ($t$) core. Using $^3$He/$t$ as a reference helps reduce the influence of the different core abundances associated with the neutron--proton asymmetry of the Au+Au system. If the effective coalescence probabilities of the two mirror channels are comparable, the \Hhe4/\Hh4 ratio is expected to approximately follow the $^3$He/$t$ ratio. A significant deviation between the two ratios could indicate different effective coalescence dynamics of the two $\Lambda$--core systems, possibly reflecting differences in $^3$He--$\Lambda$ and $t$--$\Lambda$ interactions, which may depend on the medium density~\cite{Gerstung2020}. Thus, comparing the \Hhe4/\Hh4 and $^3$He/$t$ yield ratios in the baryon-rich collision environment may help compare the in-medium $^3$He--$\Lambda$ and $t$--$\Lambda$ interactions. With a theoretical treatment of medium effects and their connection to the underlying charge-dependent $\Lambda N$ interactions, the ratio measurement could provide complementary input to studies of charge symmetry breaking (CSB) in the \Hhe4--\Hh4 hypernuclei~\cite{Haidenbauer2021}. We note that the deviation between the two ratios may also arise from other factors, such as differences in the constituent phase-space distributions at freeze-out~\cite{9m2b-99x4}.

In a thermal statistical description, the two ratios are also expected to be comparable because the baryon and strangeness chemical potentials cancel within each ratio, while the identical one-unit charge difference gives rise to the same charge-chemical-potential dependence, apart from small corrections associated with their mass and binding-energy differences.

\begin{figure}[htbp]
\centering
{\includegraphics[width=0.7\textwidth]{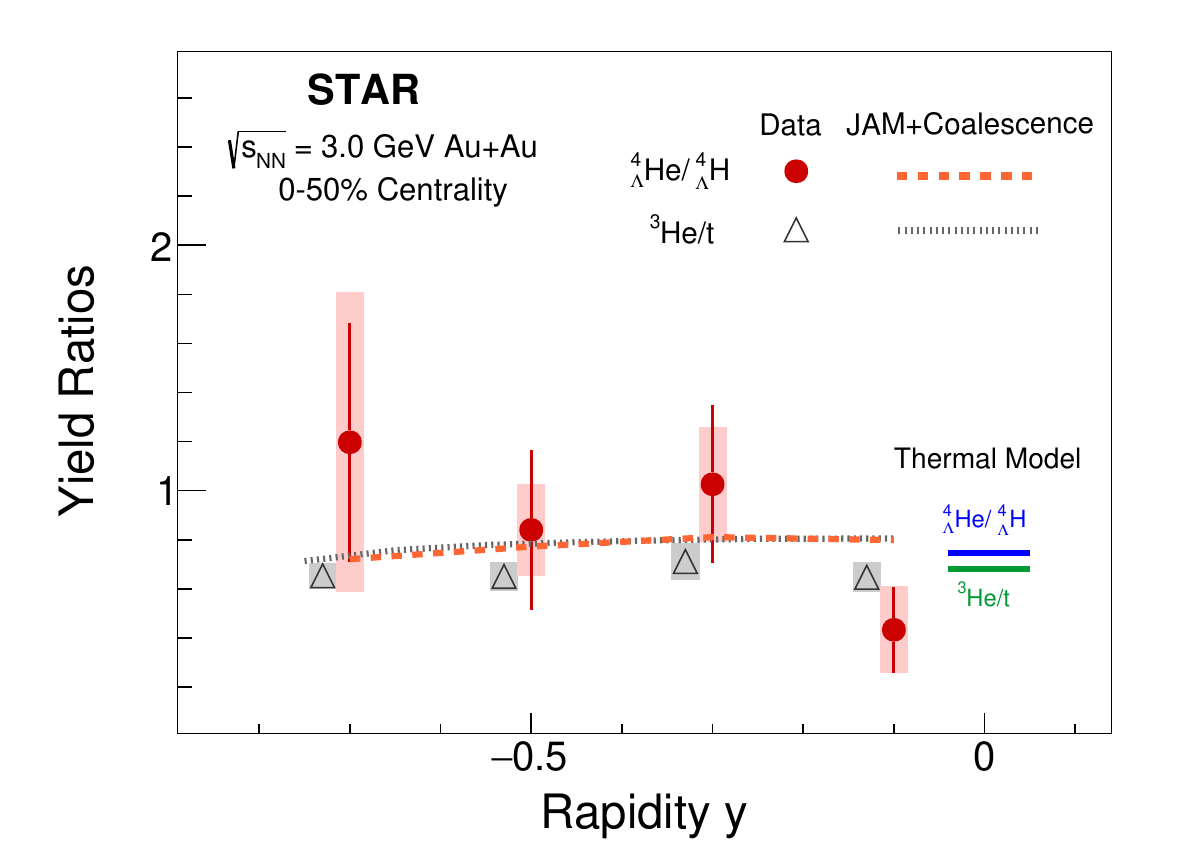}}
\caption{Rapidity dependence of the \Hhe4/\Hh4 (filled circles) and $^3$He/t (open triangles) yield ratios in 0--50\% central Au+Au collisions at $\sqrt{s_{\rm{NN}}}$ = 3.0 GeV. Vertical bars and shaded boxes denote statistical and systematic uncertainties, respectively. The orange dashed and gray dotted lines show JAM plus coalescence calculations for \Hhe4/\Hh4 and $^3$He/t, respectively; the blue and green solid lines show the corresponding Thermal-FIST predictions.}
\label{fig_result}
\end{figure}

Figure \ref{fig_result} shows the rapidity dependence of the yield ratios \Hhe4/\HF \ (filled circles) and $^3$He/$t$ (open triangles)~\cite{PhysRevC.110.054911} from 0--50\% $\sqrt{s_{\rm{NN}}}$ = 3.0 GeV Au+Au collisions. The $^3$He/$t$ yield ratio exhibits an approximately flat behavior from midrapidity toward the target-rapidity region, and the \Hhe4/\HF \ yield ratio agrees with the $^3$He/$t$ ratio as a function of rapidity. In the simple coalescence picture discussed above, this observation provides no significant evidence for a difference between the effective $^{3}\mathrm{He}-\Lambda$ and $t-\Lambda$ coalescence probabilities. The \Hhe4/\HF \ and $^3$He/$t$ ratios predicted by the JAM model plus coalescence afterburner are shown as the dashed and dotted lines in Fig.~\ref{fig_result}, respectively. The corresponding thermal-model expectations are shown by the blue and green bands, respectively. Both model calculations agree reasonably well with the measurements. We note that the absolute hypernuclear yields predicted by the thermal model remain sensitive to the overall source volume, whose determination can be further constrained by a global fit to a more complete set of hadron yields. However, the common volume dependence largely cancels in the yield ratios considered here, making them considerably less sensitive to the overall volume normalization.

\subsection[]{Conclusion}
In summary, we report the first measurement of the \Hhe4 hypernucleus yield from $\sqrt{s_{\rm{NN}}}$ = 3 GeV 0--50\% Au+Au collisions at RHIC. The rapidity dependences of d$N$/d$y$ and $\langle p_{\rm{T}} \rangle$ are compared with those of \Hh4 from the same collision system. We find that, within statistical and systematic uncertainties, the yields of \Hhe4 at different rapidities are consistent with those of \Hh4, and the \Hhe4/\Hh4 yield ratios are consistent with those of $^3$He/$t$. Using a common set of coalescence parameters for both $^3$He-$\Lambda$ and $t$-$\Lambda$, calculations from the JAM transport model with a coalescence afterburner can reproduce the yield and mean $p_{\rm{T}}$ distributions for both \Hhe4 and \Hh4, as well as the \Hhe4/\Hh4 and $^3$He/$t$ yield ratios. This suggests that their production is dominated by the coalescence mechanism. Thermal-FIST predictions with $T_{\rm{ch}}=\SI{85}{MeV}$ and $\mu_{\rm{B}}=\SI{728}{MeV}$ can also describe both ratios, but they overestimate the absolute hypernuclei yields. Future, more precise measurements may help further test the coalescence and thermal descriptions of hypernuclei production.

\section*{Acknowledgements}
We thank the RHIC Operations Group and SDCC at BNL, the NERSC Center at LBNL, and the Open Science Grid Consortium for providing resources and support. This work was supported in part by the Office of Nuclear Physics within the U.S. DOE Office of Science; the U.S. National Science Foundation; the National Natural Science Foundation of China; the Chinese Academy of Sciences; the Ministry of Science and Technology of China; the Chinese Ministry of Education; NSTC Taipei; the National Research Foundation of Korea; the Czech Science Foundation; the Ministry of Education, Youth and Sports of the Czech Republic; the Hungarian National Research, Development and Innovation Office; the New National Excellence Programme of the Hungarian Ministry of Human Capacities; the Department of Atomic Energy and the Department of Science and Technology of the Government of India; the National Science Centre and WUT ID-UB of Poland; the Ministry of Science; the German Bundesministerium f\"ur Bildung, Wissenschaft, Forschung und Technologie (BMBF); the Helmholtz Association; the Ministry of Education, Culture, Sports, Science, and Technology (MEXT); and the Japan Society for the Promotion of Science (JSPS).

\printbibliography
\end{document}